\RequirePackage{snapshot}
\documentclass[a4paper,11pt]{article}
\usepackage{pos}
\usepackage{graphicx}
\usepackage{amsmath}
\usepackage{wrapfig}
\usepackage[caption=false]{subfig}
\usepackage{siunitx}

\graphicspath{{./Figs/}}
\DeclareMathOperator{\Tr}{Tr}
\title{Centre Vortices in the Presence of Dynamical Fermions}

\FullConference{%
 The 38th International Symposium on Lattice Field Theory, LATTICE2021
  26th-30th July, 2021
  Zoom/Gather@Massachusetts Institute of Technology
}

\author{James Biddle}
\author{Waseem Kamleh}
\author{Derek Leinweber}

\emailAdd{james.biddle@adelaide.edu.au}
\emailAdd{waseem.kamleh@adelaide.edu.au}
\emailAdd{derek.leinweber@adelaide.edu.au}

\affiliation{Centre for the Subatomic Structure of Matter, Department of Physics, The
  University of Adelaide, SA 5005, Australia}

\abstract{The behaviour of centre vortices in the presence of dynamical fermions is studied for the first time in the context of the static quark potential and the Landau-gauge gluon propagator. These results indicate that in the presence of dynamical fermions, centre vortices are able to better encapsulate the long-range behaviour of QCD, compared to previous studies on pure Yang-Mills lattices. This work provides strong evidence that centre vortices are responsible for the long-range linear potential and the enhanced infrared gluon propagator, which are indicative of confinement in QCD.}
\begin{document}
\maketitle
\section{Introduction}
There is now an extensive body of lattice QCD results showing that centre vortices~\cite{tHooft:1977nqb,tHooft:1979rtg} play an important role in the confinement of quarks. Previous work has demonstrated that removal of centre vortices results in the loss of dynamical mass generation and restoration of chiral symmetry~\cite{Trewartha:2015nna,Trewartha:2017ive}, removal of the string tension~\cite{Langfeld:2003ev,Bowman:2010zr} and suppression of the infrared Landau-gauge gluon propagator~\cite{Langfeld:2001cz,Bowman:2010zr,Biddle:2018dtc}. It has also been shown that vortices alone can reproduce a linear static quark potential~\cite{Langfeld:2003ev,OCais:2008kqh,Trewartha:2015ida} and the infrared enhancement of the Landau-gauge gluon propagator. These results speak to the deep ties between centre vortices and confinement. However, the value of the vortex-only string tension obtained from pure Yang-Mills lattice studies has been consistently shown to be about $\sim 62\%$ of the full string tension. Additionally, upon removal of centre vortices the gluon propagator shows persistent infrared enhancement~\cite{Biddle:2018dtc}, indicating that there are still residual non-perturbative effects present.\\

In this work, we perform the first calculation exploring the impact of dynamical fermions on centre vortices as they govern the static quark potential and the Landau-gauge gluon propagator. By isolating the centre-vortex contributions to these quantities, we investigate how dynamical fermions impact the centre vortex model ground-state field structure. This calculation reveals significant new insight into the role of centre vortices in our understanding of QCD vacuum structure.

\section{Method}
The centre vortex picture describes the QCD vacuum via the group centre of $SU(3)$, given by the cube roots of unity. Within this picture, centre vortices appear in four dimensions as closed sheets percolating the vacuum. As the sheets pierce a three dimensional slice of the lattice, they create a vortex line. Visualisations of centre vortices on the lattice are presented in Ref.~\cite{Biddle:2019gke}. The key property is that as a vortex line pierces a planar Wilson loop, it contributes a non-trivial centre phase to the loop, namely one of $z = \exp \left( \pm\frac{ 2\pi i }{3} \right)$. From this property, it is possible to identify centre vortices on the lattice~\cite{Langfeld:2003ev,Trewartha:2015ida}. This identification procedure leads us to define three `vortex-modified' ensembles:
\begin{enumerate}
\item Original, untouched (UT) fields, $U_{\mu}(x)$,
\item Vortex-only (VO) fields, $Z_{\mu}(x)~\in~\exp \left(\frac{ 2\pi i }{3}m(x) \right),~m(x)\in \left\{ -1,\,0,\,+1 \right\}$,
\item Vortex-removed (VR) fields, $Z_{\mu}(x)^{\dagger}\, U_{\mu}(x) = R_{\mu}(x)$,
\end{enumerate}

This procedure is performed on three gauge field ensembles, each containing 200 configurations. Two of these are $(2 + 1)$ flavour dynamical ensembles from the PACS-CS collaboration~\cite{Aoki:2008sm}. The remaining ensemble is pure-gauge with a lattice spacing tuned to be similar to that of the PACS-CS ensembles. The pure-gauge ensemble provides a point of reference with which we can evaluate the impact of dynamical fermions on centre-vortex phenomenology.\\

\section{Static Quark Potential}
The static quark potential is obtained by considering the expectation value of Wilson loops, $\left\langle W(r, t) \right\rangle$, with spatial separation $r$ and temporal extent $t$~\cite{Bonnet:1999gt},
\begin{equation}
  \label{eq:1}
\langle W(r, \, t) \rangle = \sum_\alpha \lambda^{\alpha}\!(r)\, \exp \left(-V^{\alpha}\!(r)\, t\right)\, .
\end{equation}
The relevant static quark potential is given by the lowest $\alpha = 0$ state, which we extract via a variational analysis.\\

We expect the static quark potential for the original configurations to follow a Cornell potential, such that
\begin{equation}
  \label{eq:2}
  V(r) = V_0 - \frac{\alpha}{r} + \sigma\, r
\end{equation}
The Cornell potential naturally decomposes into a Coulomb term and a linear term, with the former dominating at short distances and the latter at large distances. As we expect that centre vortices will encapsulate the non-perturbative long-range physics, the vortex-only results should give rise to a linear potential~\cite{Greensite:2016pfc,DelDebbio:1998luz,Dosch:1988ha}, whereas the vortex-removed results should capture the short-range behaviour. Initially we calculate the static quark potential on the pure-gauge ensemble, making use of our three vortex-modified ensembles. To analyse the linearity of the potential at large distances we also plot a sliding linear fit to the potential with extent $-\frac{3}{2}a \leq r \leq \frac{3}{2}a$. These results are shown in Fig.~\ref{fig:sqp-PG}.\\

\begin{figure}
  \centering
  \label{fig:sqp-PG}
  \includegraphics[width=0.8\textwidth]{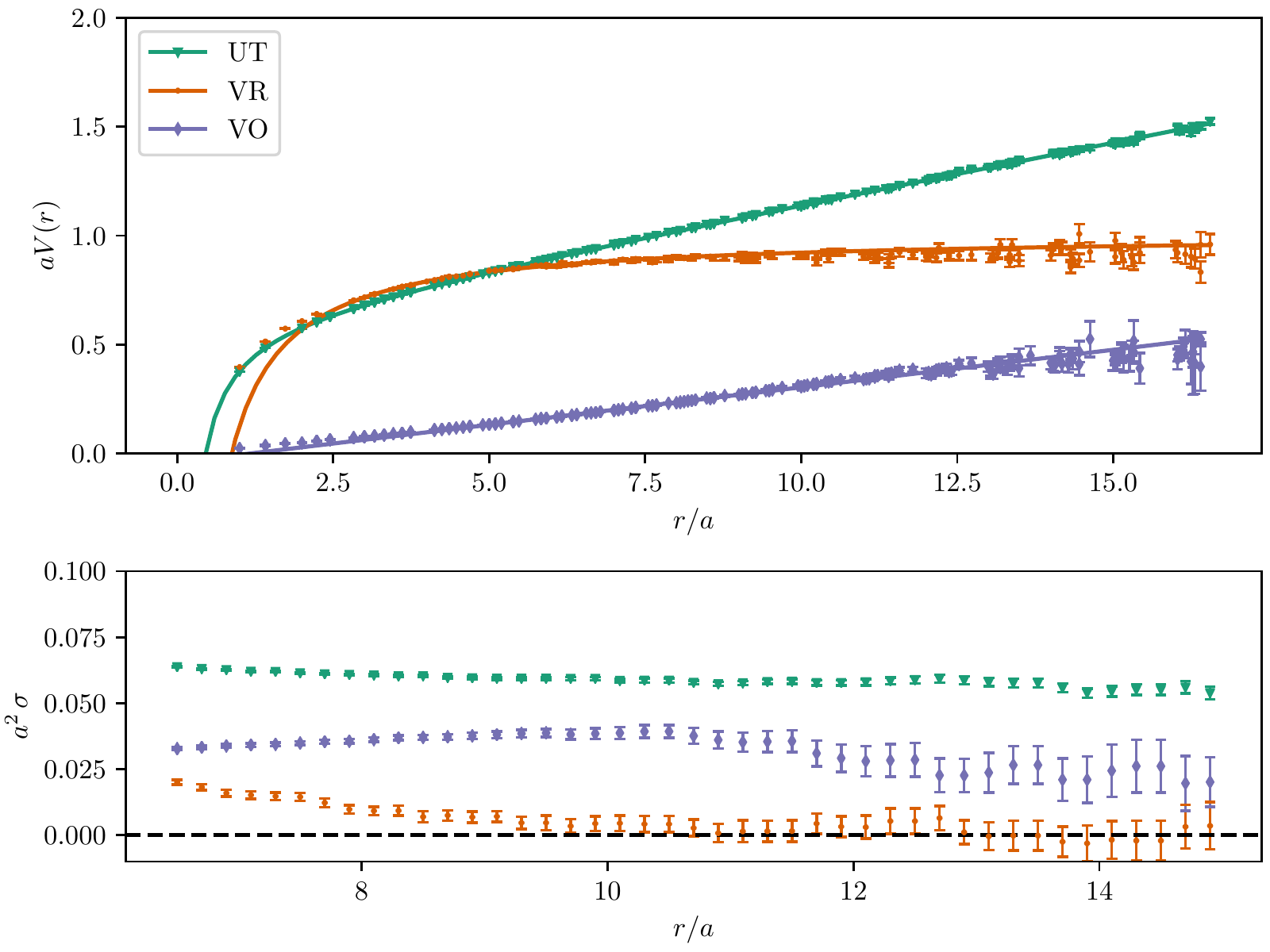}
  \caption{\label{fig:sqp-PG} The static quark potential as calculated on the
    vortex-modified pure-gauge ensembles. The lower plot shows the slope of a
    sliding linear window, as described in the text.}
\end{figure}

We observe that qualitatively the centre vortices account for the long-distance physics, as can be seen in the linearity of the vortex-only data. Furthermore, removal of vortices accounts for a complete removal of long-range potential, as visualised by the sliding fits dropping to 0. However, the value of the string tension between the original and vortex-only data differs. As has been observed previously, the vortex-only result sits at around $62\%$ of the original string tension.\\

We now introduce dynamical fermions with a pion mass of $m_{\pi} = 701\si{MeV}$ and perform the same analysis. The results of this are shown in Fig.~\ref{fig:sqp-heavy}. Here we observe a slight suppression of the original string tension, as expected due to screening effects brought about by the introduction of dynamical fermions. Most importantly, the vortex-only string tension is now in excellent agreement with the original ensemble, as can be clearly seen on the sliding fit plot. The fact that this improvement is observed even at an unphysically large pion mass is interesting, as it speaks to the dramatic effect of dynamical fermions on the vortex structure even when the observed screening effects are minimal. The effect of vortex removal is to remove confinement, with the sliding average in excellent agreement with 0 at large distances. Similar results are observed for the lightest pion mass available in the PACS-CS ensembles at $m_{\pi} = 156\si{MeV}$.\\

\begin{figure}
  \centering
  \label{fig:sqp-heavy}
  \includegraphics[width=0.8\textwidth]{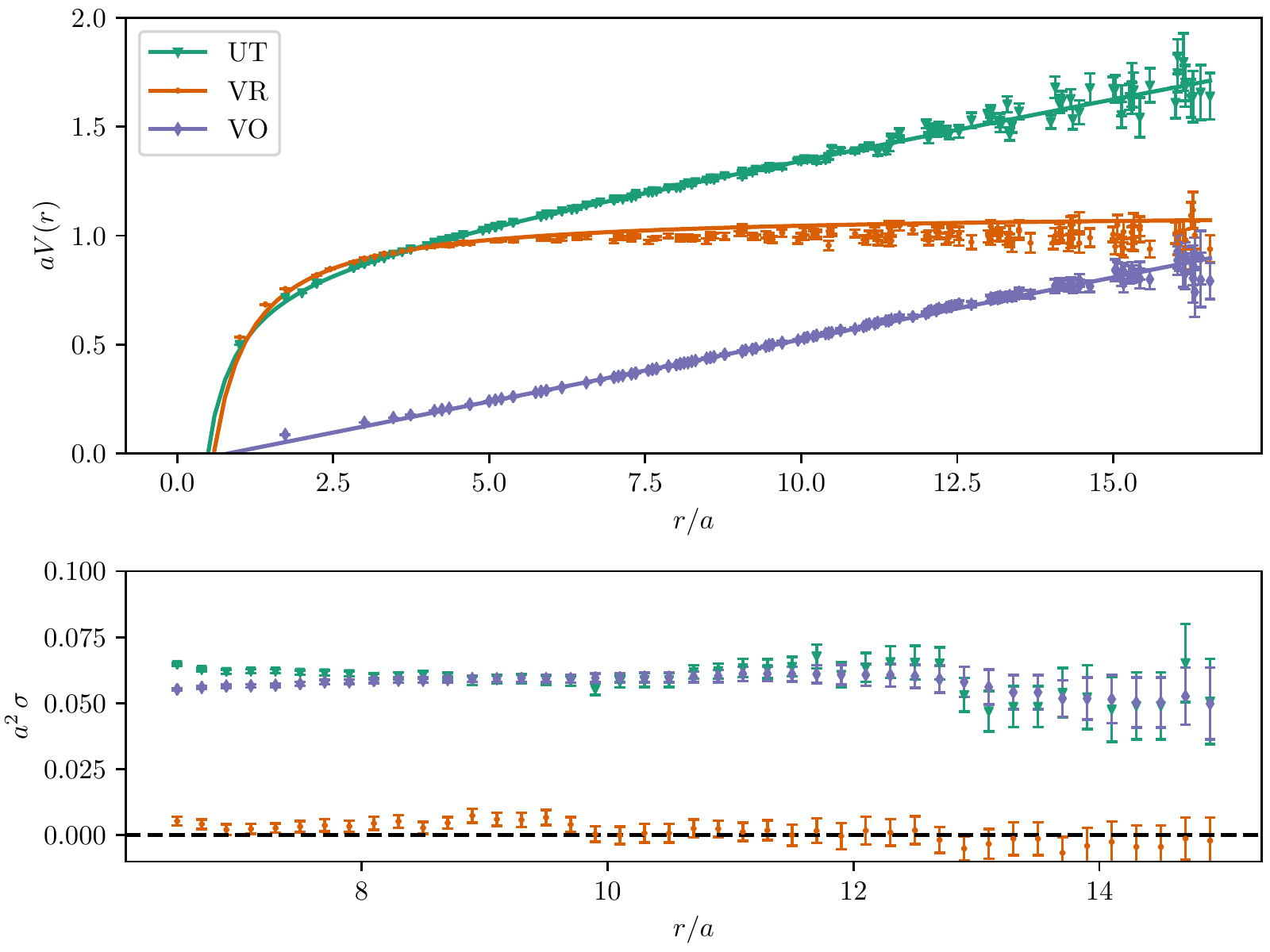}
  \caption{\label{fig:sqp-heavy} The static quark potential as calculated on the
    vortex-modified $m_{\pi} = 701\si{MeV}$ ensembles.}
\end{figure}

Overall, the introduction of dynamical fermions both improves the reproduction of the string tension from vortex-only ensembles and improves the removal of long-range potential from vortex-removed ensembles. These results are surprising, and demand further investigation into the interplay between centre vortices and dynamical fermions.

\section{Gluon Propagator}
We now turn our attention to the Landau-gauge gluon propagator in momentum space, defined on the lattice as
\begin{equation}
  \label{eq:3}
  D(q^2) =\frac{2}{3\, (n_c^2 - 1)\, V} \left\langle \Tr A_{\mu}(q)\, A_{\mu}(-q) \right\rangle\,,
\end{equation}
where $V$ is the lattice volume and $n_c = 3$ is the number of colours. The nonperturbative gluon propagator can be written as
\begin{equation}
  \label{eq:4}
  D(q^2) = \frac{Z(q^2)}{q^2}\,,
\end{equation}
where $Z(q^2)$ is the renormalisation function. In the limit as $q\rightarrow \infty$, we expect $D(q^2)\rightarrow \frac{1}{q^2}$, and as such we plot $Z(q^2)$. $Z(q^{2})$ is renormalised by setting $Z(\mu^2) = 1$ at $\mu = 5.5 \si{GeV}$.\\

\begin{figure}
  \centering
  \includegraphics[height=0.6\textwidth]{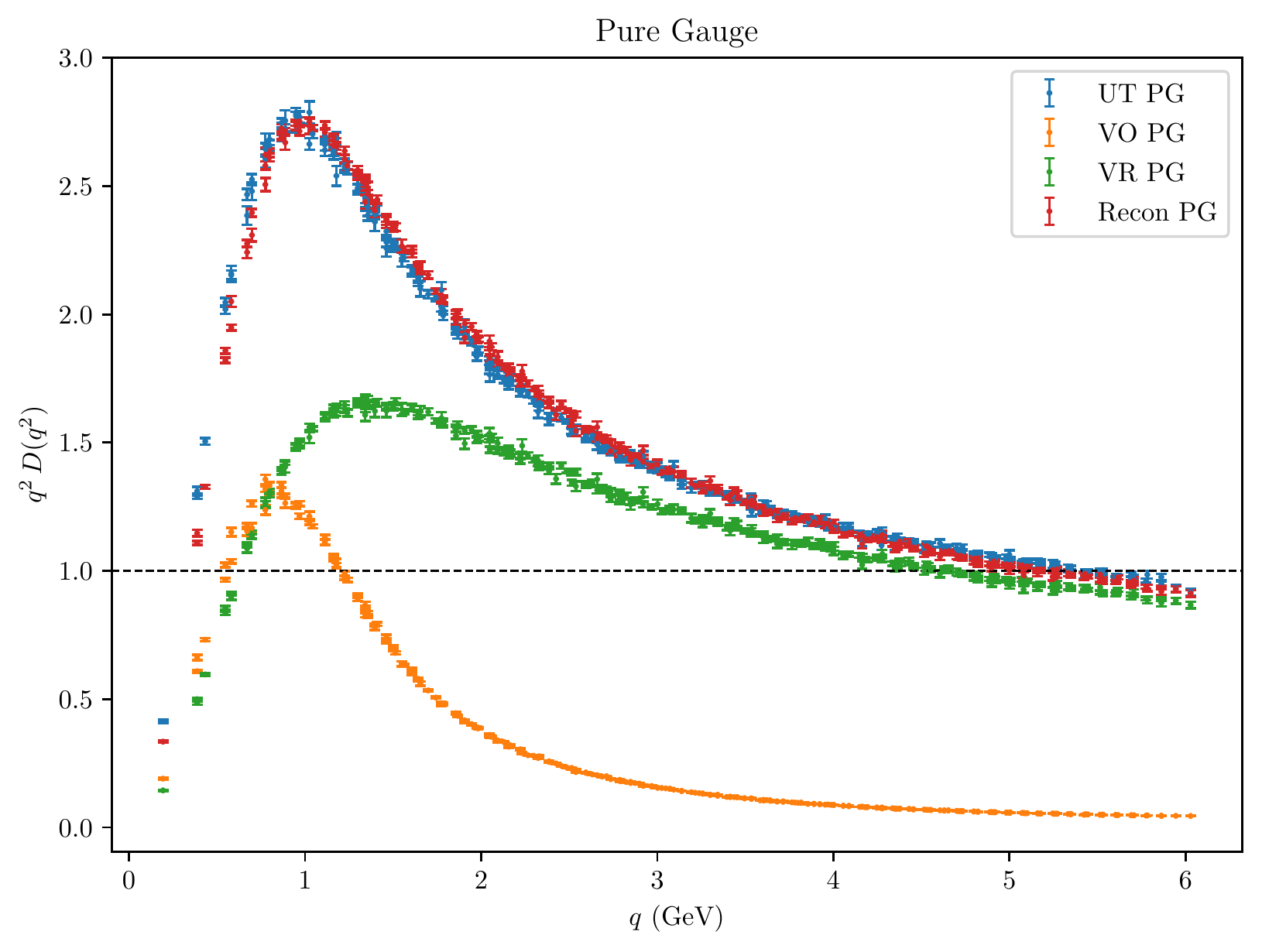}
  \caption{\label{fig:gprop-PG} Pure-gauge vortex-modified gluon propagators.}
\end{figure}

Beginning with the pure-gauge ensemble, the vortex-modified propagators are as shown in Fig.~\ref{fig:gprop-PG}. The original propagator is epitomised by infrared enhancement and a trend towards unity at large momentum. Here we see that vortex removal suppresses the infrared enhancement, but retains the high momentum behaviour. Conversely, the vortex-only propagator contains only infrared strength, with a complete removal of high momentum behaviour. We also consider reconstructing the original propagator as a linear combination of the vortex-only and vortex-removed propagators. Here we see that we can achieve a full reproduction of the original propagator via this method, indicating that vortices effectively partition the propagator short and long-range information. However, there is still substantial residual infrared strength in the vortex-removed propagator.\\

\begin{figure}
  \centering
  \includegraphics[height=0.6\textwidth]{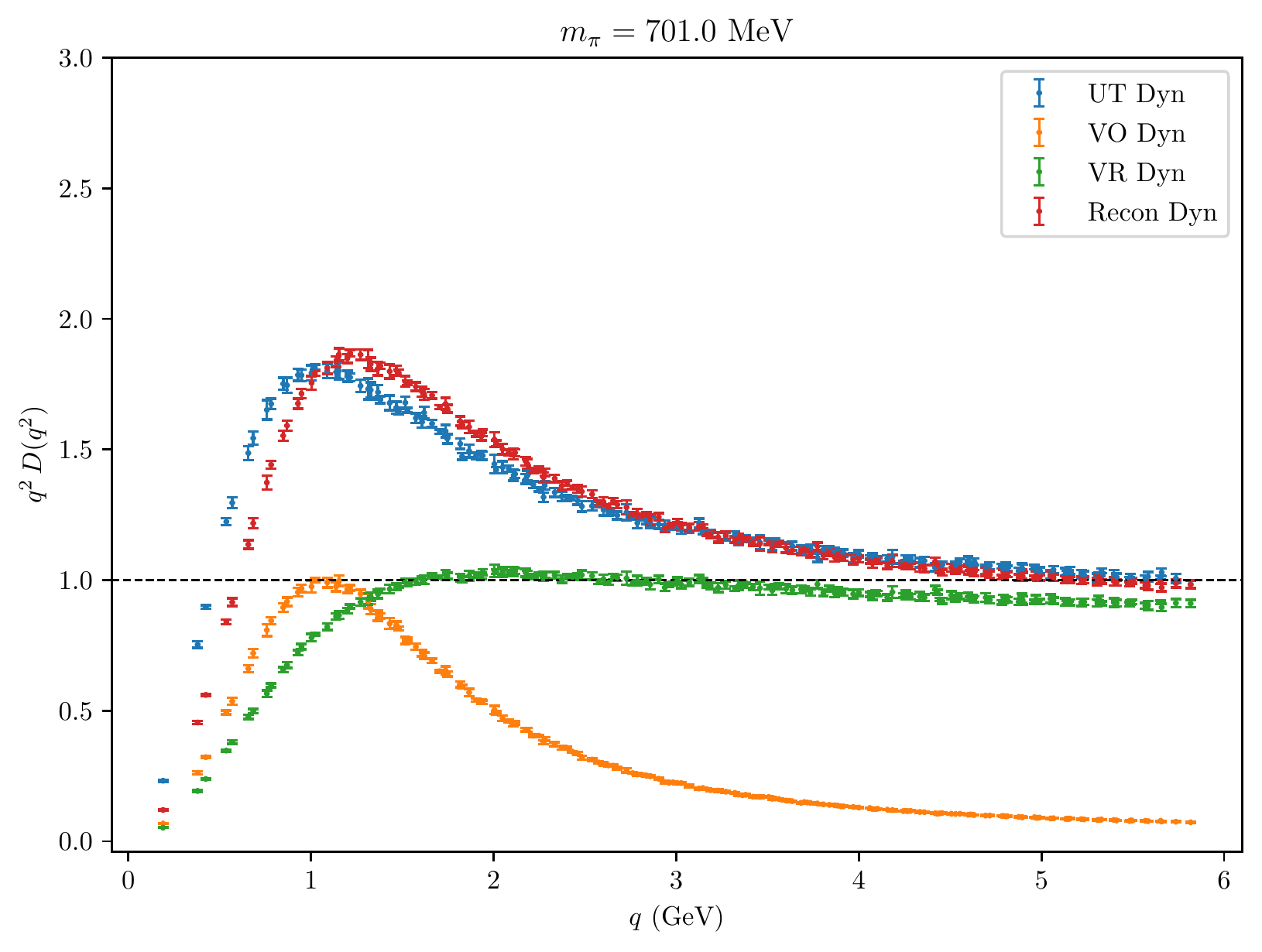}
  \caption{\label{fig:gprop-heavy} $m_{\pi} = 701\si{MeV}$ vortex-modified gluon propagators.}
\end{figure}

Turning to the dynamical ensemble with $m_{\pi} = 701\si{MeV}$, the vortex-modified propagators now appear as in Fig.~\ref{fig:gprop-heavy}. We see a noticeable screening of the propagator, coupled with improvement in the ability of vortex-removal to remove infrared enhancement. The ability to reconstruct the original propagator from its vortex components is diminished, and at present the origins of this discrepancy are under investigation. However, the increased suppression of infrared strength after vortex removal is an important point, as it indicates a substantial enhancement in the isolation of infrared strength into the vortex-only propagator.\\

\section{Conclusion}
We have presented the first study of the impact of dynamical fermions on centre vortices in the context of the static quark potential and the gluon propagator. In regards to the static quark potential, we find that centre vortices can quantitatively capture the string tension, and that vortex removal eliminates the long distance potential. This is in stark contrast to the pure-gauge situation, where vortices only qualitatively reproduce the long-range potential.\\

When considering the gluon propagator, the results are encouraging. The increased ability of vortex removal to suppress infrared enhancement is an important result that demands further investigation. The discrepancy in reconstructing the original propagator from its vortex-modified components is also being studied further.\\

These early results are exciting and perhaps unexpected. The results presented here show a significant change in the ability of centre vortices to capture the underlying physics of confinement. Further calculation is needed to clarify this picture, but so far these results point to a host of exciting relationships between centre vortices and dynamical fermions on the lattice.

\acknowledgments{We thank the PACS-CS Collaboration for making their 2 +1
  flavour configurations available via the International Lattice Data Grid
  (ILDG). This research was undertaken with the assistance of resources from the
  National Computational Infrastructure (NCI), provided through the National
  Computational Merit Allocation Scheme and supported by the Australian
  Government through Grant No. LE190100021 via the University of Adelaide
  Partner Share. This research is supported by Australian Research Council
  through Grants No. DP190102215 and DP210103706. WK is supported by the Pawsey
  Supercomputing Centre through the Pawsey Centre for Extreme Scale Readiness
  (PaCER) program.}

\bibliographystyle{JHEP}
\bibliography{ConferenceProceedings}
      %
      %
      %
      %

\end{document}